\documentclass[aps,floatfix,showpacs,pra,twocolumn]{revtex4}
\usepackage{graphicx,bm}

\begin{document}
\title{Metastability Exchange Optical
  Pumping of Helium-3 at High Pressures and 1.5~T:
Comparison of two Optical Pumping Transitions}
\author{M. Abboud}
\email{marie.abboud@lkb.ens.fr}
\author{A. Sinatra}
\author{G. Tastevin}
\author{P.-J. Nacher}
\affiliation {Laboratoire Kastler Brossel, Ecole Normale
  Sup\'{e}rieure, 24 rue Lhomond, 75005 Paris, France \footnote
{Laboratoire Kastler Brossel is a unit\'e de recherche de l'Ecole
Normale Sup\'erieure et de l'Universit\'e Pierre et Marie Curie,
associ\'ee au CNRS (UMR 8552).}}
\author{X. Ma\^{\i}tre}
\affiliation {U2R2M, Universit\'e Paris-Sud and CIERM, H\^opital de Bic\^etre,
94275 Le Kremlin-Bic\^etre Cedex, France \footnote {U2R2M (Unit\'e de
Recherche en R\'esonance Magn\'etique M\'edicale) is a unit\'e de
recherche de l'Universit\'e Paris-Sud, associ\'ee au CNRS (UMR
8081).}}

\begin{abstract}
{\bf Abstract:} \\
At low magnetic field, metastability exchange optical pumping of
helium-3 is known to provide high nuclear polarizations for pressures
around 1~mbar. In a recent paper, we demonstrated that
operating at 1.5~T can significantly improve the results of
metastability exchange optical pumping at high pressures. Here, we compare the
performances of two different optical pumping lines at 1.5~T, and show
that either the achieved nuclear polarization or the production rate
can be optimized.
\end{abstract}

\maketitle

\section{Introduction}

\label{intro} Highly polarized helium-3 is used in various fields of science,
for example, to perform magnetic resonance imaging (MRI) of air spaces in
human lungs \cite{albert,moller}, or to prepare spin filters for neutrons
\cite{becker} and polarized targets for nuclear physics \cite{xu}. The most
successful methods presently used to polarize helium-3 are spin-exchange
optical pumping using alkali atoms \cite{bouchiat,happer}, and pure-helium
metastability exchange optical pumping \cite{colegrove,nacher85}. The
applications have driven research towards improvement in terms of photon
efficiency, steady-state polarization, and production rate, both for spin
exchange optical pumping \cite{babcock}, and metastability exchange optical
pumping \cite{becker,gentile}. The metastability exchange technique was
demonstrated by Colegrove, Schearer, and Walters over forty years ago
\cite{colegrove}. In standard conditions, metastability exchange optical
pumping is performed at low pressure (1~mbar) in a guiding magnetic field up
to a few mT. Metastable $2^{3}$S-state atoms are produced using a
radiofrequency discharge. They are optically pumped using the $2^{3}$S-$2^{3}%
$P transition at 1083~nm. The electronic polarization is transferred to the
nuclei by hyperfine interaction. Through metastability exchange collisions,
nuclear polarization is transferred to ground state helium-3 atoms.
Metastability exchange optical pumping in standard conditions provides in a
few seconds high nuclear polarizations (up to 90\% at 0.7~mbar \cite{batz}).
Unfortunately, the achieved nuclear polarization rapidly drops down when the
helium-3 pressure exceeds a few mbar \cite{gentile,cracovie}. Therefore, a
delicate polarization-preserving compression stage is necessary for all
applications needing a dense sample.

We recently demonstrated that operating at 1.5~T can significantly improve the
nuclear polarization achieved at high pressures \cite{abboud}, using one of
the most intense lines in the $2^{3}$S-$2^{3}$P absorption spectrum. Here, we
show that a different choice of optical pumping transition can further improve
the steady-state polarization although the production rate is slightly lower.
We also describe more precisely the experimental protocol, and demonstrate the
consistency of the optical absorption technique for dynamic measurement of the
nuclear polarization in presence of the optical pumping laser.

\begin{figure}[b]
\centerline{\includegraphics[width=8cm,clip=]{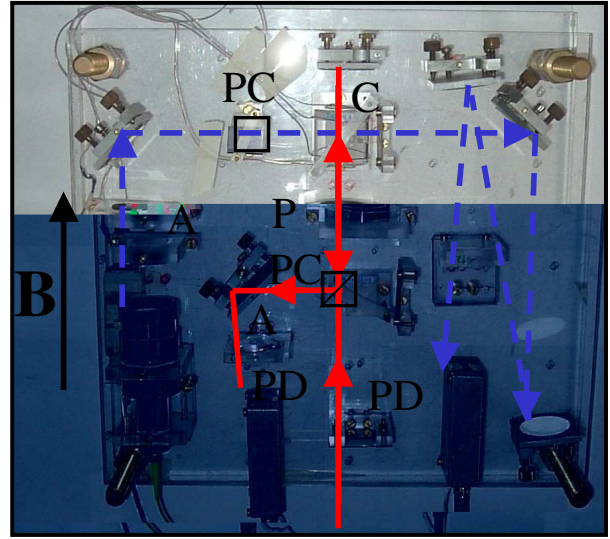}}
\caption{Picture of the experimental apparatus of optical pumping at 1.5~T,
containing the optical elements and the sealed optical pumping cell. The pump
laser beam (solid arrow) is parallel to the magnetic field \textbf{B}, and the
probe laser beam (dashed arrow) is perpendicular to the magnetic field. C:
optical pumping cell, PC: polarizing cube, P: quarter-wave plate, PD:
photodiode, A: attenuator.}%
\label{fig:setup}%
\end{figure}

\section{Experimental}

\label{experimental}

\subsection{Setup}

\label{setup} 
\begin{figure}[b]
\centerline{\includegraphics[width=7.2cm,clip=]{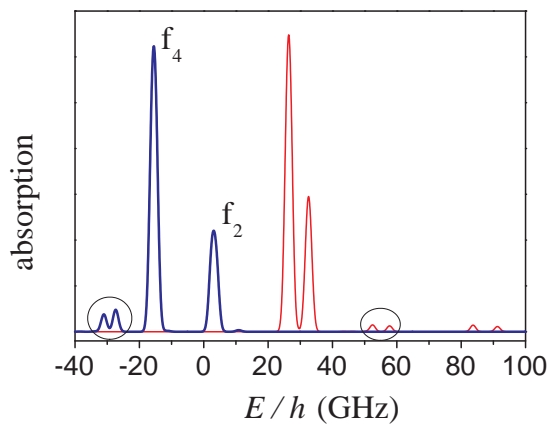}}%
\caption{Computed absorption spectra for $\sigma^{-}$ (thick line) and
$\sigma^{+}$ (thin line) light at 1.5~T. f$_{4}$ and f$_{2}$ are the two
optical pumping lines used in our experiments. The circled peaks on the left
(respectively on the right) correspond to the probe lines used when the pump
laser is tuned on f$_{4}$ (respectively on f$_{2}$). Spectral line positions
are defined as in reference \cite{courtade}.}%
\label{fig:spectre}%
\end{figure}

Experiments are performed in the bore of a clinical MRI scanner
providing a homogeneous 1.5~T magnetic field. The experimental apparatus is
shown in Fig.\ref{fig:setup}. The helium-3 gas is enclosed in a sealed
cylindrical Pyrex cell, $5$ cm in diameter and $5$ cm in length. Cells filled
with pressure $P$=1.33, 8, 32, and 67~mbar of pure helium-3 are used. A
radiofrequency high voltage applied to electrodes on the outside of the cell
generates a weak electrical discharge in the gas. It is used to populate the
$2^{3}$S state and maintain a metastable atoms density n$_{m}$ in the range
1-8$\times$10$^{10}$ atoms/cm$^{3}$, depending on the applied voltage and the
gas pressure. The optical pumping laser is a 50~mW single mode laser diode
amplified by a 0.5~W ytterbium-doped fiber amplifier \cite{chernikov}. The
laser wavelength can be tuned by temperature control over the entire spectrum
of the $2^{3}$S-$2^{3}$P transition of helium ($\sim$150~GHz at 1.5~T, see
Fig.\ref{fig:spectre}). The circular polarization of the pump beam is obtained
using a combination of polarizing cube and quarter-wave retarding plate. The
pump beam is back-reflected after a first pass in the cell to enhance its
absorption, and collected by a photodiode to measure its absorption.
The probe beam is provided by another single mode laser diode. It
is attenuated and linearly polarized perpendicularly to the magnetic field.
The absorption of probe and pump lasers are measured using a modulation
technique. The discharge intensity is modulated at $133$~Hz and the
absorptions are measured with lock-in amplifiers. The average values of the
transmitted probe and pump intensities are also recorded. Laser sources and
electronics remain several meters away from the magnet bore in a low-field region.

\subsection{Optical pumping configuration}
\label{OPconfig}
The structure of the $2^{3}$S-$2^{3}$P transition and Zeeman sublevels of
helium-3 at 1.5~T is described in \cite{courtade,abboud}.
\begin{figure}[b]
\centerline{\includegraphics[width=6cm,clip=]{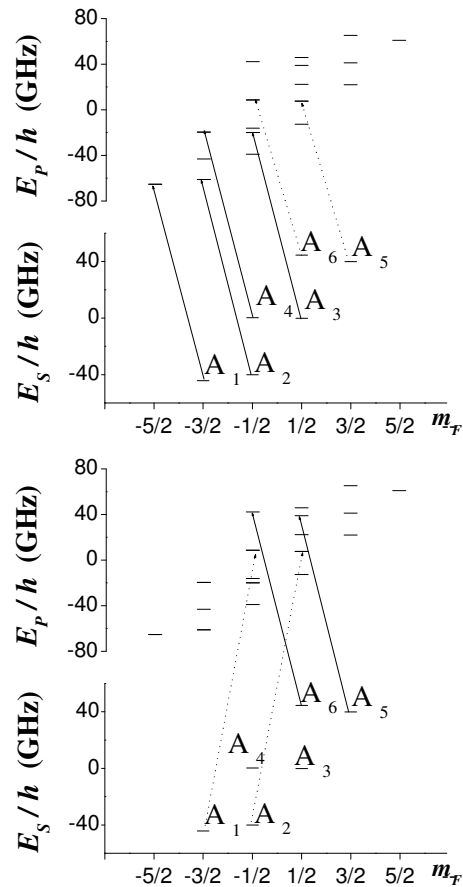}}%
\caption{Energies and magnetic quantum numbers $m_{F}$ of the helium-3
sublevels at 1.5~T for the $2^{3}$S (E$_{S}$) and $2^{3}$P (E$_{P}$) states.
The upper (lower) scheme corresponds to the f$_{4}$ (f$_{2}$) optical pumping
configuration. The $\sigma^{-}$ pumping transitions (solid lines) are
displayed. The populations of the sublevels A$_{5}$ and A$_{6}$ (A$_{1}$ and
A$_{2}$) not addressed by the f$_{4}$ (f$_{2}$) optical pumping transition are
measured using the $\sigma^{-}$ ($\sigma^{+}$) probe transitions (dashed
lines). Level names A$_{1}$ to A$_{6}$ and energy zeroes are defined as in
reference \cite{courtade}.}%
\label{fig:OPscheme}%
\end{figure}
In our experiments, optical pumping is performed using one of the
two $\sigma^{-}$ pumping lines labeled f$_{4}$ or f$_{2}$ in the absorption
spectrum displayed in Fig.\ref{fig:spectre}. The f$_{4}$ line consists of four
unresolved transitions spreading over 1.31~GHz. Given the Doppler width of
$\ $helium-3 at room temperature (2~GHz~FWHM), it addresses simultaneously
four metastable sublevels A$_{1}$ to A$_{4}$ with $m_{F}$=-3/2, -1/2 and 1/2,
where $m_{F}$ is the magnetic quantum number for the total angular momentum
(see Fig.\ref{fig:OPscheme}). The f$_{2}$ line (two transitions split by
1.37~GHz) simultaneously addresses sublevels A$_{5}$ and A$_{6}$ with $m_{F}%
$=1/2 and 3/2. The population transfer into A$_{5}$ and A$_{6}$ for f$_{4}$
pumping (or into A$_{1}$ to A$_{4}$ for f$_{2}$ pumping) occurs as follows:
excitation by laser absorption, collisional redistribution in the $2^{3}$P
state, and spontaneous emission.

\subsection{Optical measurement of nuclear polarization}

\label{measurement} The optical detection method used in our experiments is
based on absorption measurements using a weak probe beam.
\begin{figure}[b]
\centerline{\includegraphics[width=7.0cm,clip=]{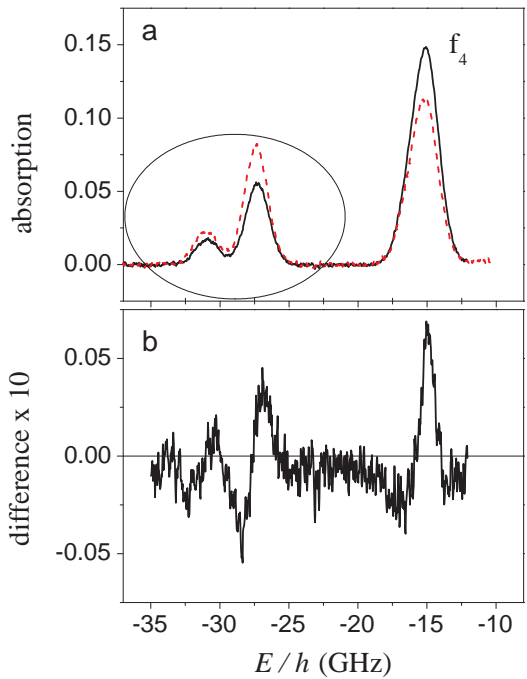}}%
\caption{\textbf{a:} Absorption signals recorded with pump laser on (dashed
line) and off (solid line). They are performed in the 8~mbar cell with 0.25~W
pump laser power for f$_{4}$ pumping after reaching $M_{eq}=0.43$. The circled
peaks involving sublevels A$_{5}$ and A$_{6}$, not addressed by the pump
laser, are used to compute $M$. \textbf{b:} The lower graph is a residue plot
showing the difference between the solid line data in Fig.\ref{fig:spintemp}a
and a computed spin temperature distribution spectrum.}%
\label{fig:spintemp}%
\end{figure}
This absorption technique does not need any calibration and can be
used at arbitrary magnetic field, and pressure \cite{courtade}. It relies on
the fact that in the absence of optical pumping, metastability exchange
collisions impose a spin temperature distribution for the metastable
populations $a_{m_{F}}\propto e^{\beta m_{F}}$, where $1/\beta$ is the spin
temperature in the $2^{3}$S state related to the nuclear polarization $M$ in
the ground state $M=(e^{\beta}-1)/(e^{\beta}+1)$.

In practice, the probe laser frequency is swept over two lines. Peaks
amplitudes are precisely measured by a fit to a Voigt absorption profile.
\begin{figure}[t]
\centerline{\includegraphics[width=7.0cm,clip=]{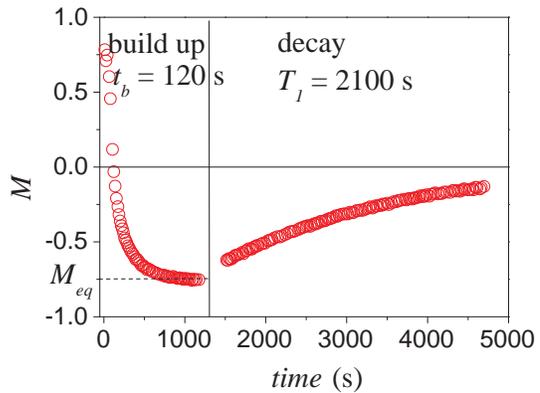}}%
\caption{Example of dynamic monitoring of the nuclear polarization $M$ in the
8~mbar cell. Starting from a gas prepared with $M>0$ at time $t=0$, the pump
laser is tuned on the f$_{2}$ line with 0.5~W power. After reaching the
steady-state nuclear polarization $M_{eq}=-0.75$, the pump laser is turned off
at $t=1300$~s, and the discharge-induced decay of the polarization is
observed.}%
\label{fig:example}%
\end{figure}
The population ratio of the two hyperfine sublevels addressed by
the probe lines is then found using the field-dependent computed transition
probabilities, and is used to calculate the spin temperature. Hence, the
ground-state nuclear polarization $M$ is inferred \cite{courtade}.

Polarization build-up is monitored in the presence of the optical pumping
beam. Therefore, the probe absorption measurements must involve metastable
sublevels not addressed by the optical pumping laser. In our configuration
(see Fig.\ref{fig:OPscheme}), the populations of sublevels A$_{5}$ and A$_{6}$
(respectively A$_{1}$ and A$_{2}$) are measured for f$_{4}$ (respectively
f$_{2}$) pumping. Fig.\ref{fig:spintemp}a shows typical absorption spectra
recorded in the absence and in the presence of the pump laser. In the absence
of the pump laser (solid line), the spectrum is accurately fit by the spectrum
computed assuming a spin temperature distribution (residue plot in
Fig.\ref{fig:spintemp}b). The presence of the pump strongly affects the
population distribution, with efficient population transfer from the pumped
levels to A$_{5}$ and A$_{6}$ and modifies the absorption profile (dashed
line). However the ratio of populations in sublevels A$_{5}$ and A$_{6}$
remains unaffected, and an absorption measurement still accurately provides
the correct value for $M.$

An example of dynamical measurement of polarization build-up and decay is
shown in Fig.\ref{fig:example} for the 8~mbar cell. Several scans of the probe
laser are recorded successively, and the value of the nuclear polarization $M$
is inferred as a function of time. The gas is initially polarized to the
equilibrium positive value of $M$ achieved with f$_{4}$ pumping. At time
$t=0$, the pump laser is tuned to the f$_{2}$ line. When the new steady-state
polarization $M_{eq}$ is reached, the pump laser is turned off and
polarization decays by discharge-induced relaxation.

\section{Results}
\label{results}
\begin{figure}[h]
\centerline{\includegraphics[width=7.0cm,clip=]{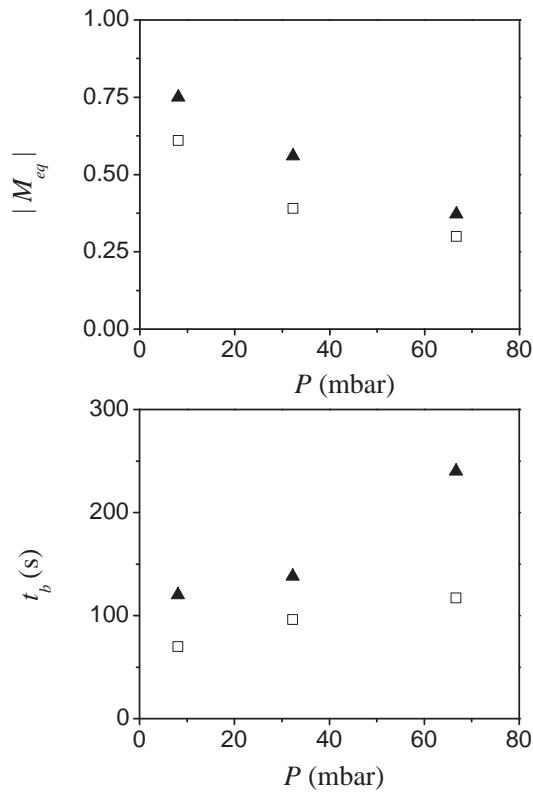}}%
\caption{Absolute values of steady-state nuclear polarizations $M_{eq}$ and
polarization build-up times $t_{b}$ as a function of the helium-3 pressure at
1.5~T. Triangles (respectively squares) are data obtained when the pump laser
(0.5~W) is tuned to f$_{2}$ (respectively f$_{4}$). The measured $n_{m}$ and
$T_{1}$ are: at 8~mbar, 3.54$\times$10$^{10}$ at./cm$^{3}$ and 2100~s, at
32~mbar, 2.47$\times$10$^{10}$ at./cm$^{3}$ and 1490~s, at 67~mbar,
1.30$\times$10$^{10}$ at./cm$^{3}$ and 1190~s.}%
\label{fig:bestMtb}%
\end{figure}
Several experimental parameters influence the performances of
optical pumping: the cell geometry, the gas pressure, the discharge conditions
(voltage, electrodes configuration) which impact on $T_{1}$ and $n_{m}$, the
optical pumping transition, and the pump laser power. A systematic study of
optical pumping using the f$_{4}$ line has been reported in \cite{abboud}.
Here, we focus on the comparison of optical pumping with f$_{4}$ and f$_{2}$
lines in the same cells. We present selected results obtained with identical
discharge conditions for f$_{4}$ and f$_{2}$, using a 0.5~W pump power.

The f$_{2}$ line yields higher steady-state polarizations but f$_{4}$ allows
pumping with significantly shorter build-up times (Fig.\ref{fig:bestMtb}). As
a result, the magnetization production rates $R_{a}=P\times M_{eq}/t_{b}$ are
slightly lower for f$_{2}$ (see Table~\ref{table:Ra}). The main difference
between the two optical pumping lines actually lies in the photon efficiency,
defined as the number of polarized nuclei per photon absorbed by the gas. A
simple calculation, based on the average angular momentum transfer from
polarized light to atoms during one absorption-collisional
redistribution-spontaneous emission cycle, shows that the photon efficiency of
f$_{2}$ should be approximately twice that of f$_{4}$. This has been
experimentally checked, at various laser powers. Magnetization production
rates for f$_{2}$ pumping are yet slightly lower due to the lower absorption
of the pump beam on this line with respect to f$_{4}$.
\begin{table}[h]
\caption{Magnetization production rates $R_{a}=P\times M_{eq}/t_{b}$ versus
gas pressure (data in Fig.\ref{fig:bestMtb}).}
\begin{center}%
\begin{tabular}
[c]{c|c|c|}\cline{2-3}%
& \multicolumn{2}{|c|}{$R_{a}$ (mbar/s)}\\\hline
\multicolumn{1}{|c|}{$P$ (mbar)} & f$_{4}$ & f$_{2}$\\\hline
\multicolumn{1}{|c|}{8} & 0.07 & 0.05\\\hline
\multicolumn{1}{|c|}{32} & 0.13 & 0.12\\\hline
\multicolumn{1}{|c|}{67} & 0.17 & 0.11\\\hline
\end{tabular}
\end{center}
\label{table:Ra}%
\end{table}

\section{Discussion}

We have compared the performances of two different optical pumping lines at
1.5~T in helium-3 gas, at high pressures (up to 67~mbar). The strongest line
of the $\sigma^{-}$ absorption spectrum (f$_{4}$) yields the highest
magnetization production rates. However, the highest steady-state nuclear
polarizations (up to $M_{eq}=-0.75$ at $8$~mbar) are achieved using a weaker
line with higher photon efficiency. f$_{2}$ pumping, requiring fewer absorbed
photons to achieve the same production rate, is thus advantageous when long,
optically thick pumping cells are used.

Given the structure of sublevels and optical transitions, one could expect
that the two most intense $\sigma^{+}$ lines would be just as efficient as the
corresponding $\sigma^{-}$ lines. Similar optical pumping performances have
indeed been obtained, although no systematic comparison has been carried out.

An analysis of all optical pumping data collected at 1.5~T is under way using
a detailed model of metastability exchange optical pumping in high field
conditions, and will be the subject of a forthcoming paper.

\end{document}